\documentstyle[12pt,amsmath,latexsym]{article}
\def\Box{\hbox{$\sqcup$\kern-0.66em\lower0.03ex\hbox{$\sqcap$}}}

\begin{document}
\begin{titlepage}
{~}
\vskip 3truecm 
\begin{center}
\Large\bf
Proof of Polyakov conjecture for general elliptic singularities
\footnote{This work is  supported in part by M.U.R.S.T.}
\end{center}

\vskip 1truecm
\begin{center}
{Luigi Cantini$~^{a}$, Pietro Menotti$~^{b}$, Domenico Seminara}$^{c}$\\  
\end{center}
\begin{center}
\vskip 1truecm
\small\it $^{a}$ Scuola Normale Superiore, 56100 Pisa, Italy\\
{\small\it and INFN, Sezione di Pisa}\\
e-mail: cantini@df.unipi.it\\
{\small\it $^{b}$ Dipartimento di Fisica dell'Universit{\`a}, 56100 Pisa, 
Italy}\\
{\small\it and INFN, Sezione di Pisa}\\
e-mail: menotti@df.unipi.it\\
{\small\it $^{c}$ Dipartimento di Fisica dell'Universit{\`a}, 50125
Firenze, Italy}\\
{\small\it and INFN, Sezione di Firenze}\\
e-mail: seminara@fi.infn.it\\
\end{center}
\vskip 1truecm
\end{titlepage} 

\begin{abstract}
A proof is given of Polyakov conjecture about the accessory parameters of
the $SU(1,1)$ Riemann-Hilbert problem for general elliptic
singularities on the Riemann sphere. Its relevance to $2+1$
dimensional gravity 
is stressed.
\end{abstract}

\section{Introduction}

Polyakov made the following conjecture \cite{conj} on the accessory parameters
$\beta_n$ which appear in the solution of the $SU(1,1)$
Riemann-Hilbert problem

\begin{equation}\label{conjecture}
-\frac{1}{2\pi} d S_P = \sum_n\beta_n dz_n + c.c. 
\end{equation}
where $S_P$ is the regularized Liouville action \cite{takh1},
$S_P = \lim_{\epsilon \rightarrow 0} S_\epsilon $
with \footnote{Our conventions are slightly different from those of
ref.\cite{takh1,ZT1}; our field $\phi$ is related the the field $\psi$ of
ref.\cite{ZT1} by $\phi =\psi +\ln 2$}
$$
S_\epsilon [\phi] =\frac{i}{2} \int_{X_\epsilon} (\partial_z\phi 
\partial_{\bar z} \phi +\frac{e^\phi}{2}) dz\wedge d\bar z
+\frac{i}{2}\sum_n g_n\oint_{\gamma_n}\phi(\frac{d\bar z}{\bar z -\bar
z_n}- \frac{d z}{ z - z_n})
$$
\begin{equation}\label{Sepsilon}
+\frac{i}{2}g_\infty\oint_{\gamma_\infty}\phi(\frac{d\bar z}{\bar z}- \frac{d
z}{z}) 
-\pi\sum_n g_n^2 \ln\epsilon^2 -\pi g_\infty^2\ln\epsilon^2,~~~~{\rm
where}~~~~dz\wedge d\bar z = -2i dx\wedge dy 
\end{equation}
and $X_\epsilon$ is the disk of radius $1/\epsilon$ in the complex plane from
which disks of radius $\epsilon$ around all singularities have been
removed; $\gamma_n$ are the boundaries of the small disks and
$\gamma_\infty$ is the boundary of the large disk.

 In eq.(\ref{conjecture}) $S_P$ has to be computed on the solution of the
inhomogeneous Liouville equation which arises
from the minimization of the action i.e.
\begin{equation}\label{pic}
4 \partial_z \partial_{\bar z}\phi = e^{\phi} +4\pi\sum_n g_n\delta^2(z-z_n)  
\end{equation}
with asymptotic behavior at infinity $\phi= -g_\infty \ln z\bar
z+O(1)$.  
Such a conjecture plays an important role in the quantum Liouville
theory \cite{takh2} and in the ADM formulation of $2+1$ dimensional
gravity 
\cite{MS,CMS}. The conjecture is interesting in itself as it gives
a new meaning to the rather elusive accessory parameters
\cite{hempel,yoshida} of the Riemann-Hilbert
problem. In particular it implies that the form 
$
\omega = \sum_n\beta_n dz_n + c.c.
$
is exact.

Zograf and Takhtajan \cite{ZT1} provided a proof of
eq.(\ref{conjecture}) for parabolic singularities using the technique
of mapping the quotient of the upper half-plane by a fuchsian group to
the Riemann surface and exploiting certain properties of the harmonic
Beltrami differentials. In addition they remark that the same
technique can be applied when some of the singularities are elliptic
of finite order. The case of only parabolic singularities is of
importance in the quantum Liouville theory \cite{takh2} as such
singularities 
provide the sources from which to compute the correlation
functions. On the other hand in $2+1$ gravity one is faced with
general elliptic singularities and here the mapping technique cannot
be directly applied. (In the case of elliptic singularities with
rational $g_n$ some progress in the mapping technique that are
relevant to this problem were made in \cite{kra}). 
As a matter of fact we shall see that the case of elliptic
singularities is more closely related to the theory of elliptic non
linear differential equations (potential theory) than to the theory of
fuchsian groups.

In a series of papers at the turn of the past century Picard
\cite{picard} proved that eq.(\ref{pic}) for real $\phi$
with asymptotic behavior at infinity 
\begin{equation}
\phi(z) = -g_\infty\ln(z\bar z) + O(1)
\end{equation}
and $-1<g_n,~~1<g_\infty$ (which excludes the case of punctures) and
$\sum_n g_n +g_\infty < 0$ 
admits one and only one solution (see also \cite{poincare}). Picard
\cite{picard} achieved the solution of (\ref{pic}) through an
iteration process exploiting Schwarz alternating procedure. The same
problem has been considered recently with modern variational
techniques by Troyanov \cite{troyanov}, obtaining results which
include Picard's findings. 
The interest of such results is that they solve the following variant
of the Riemann-Hilbert problem: at $z_1,\dots z_n$ we are given not
with the monodromies but with the class, characterized by $g_j$, of
the elliptic monodromies with the further request that all such
monodromies belong to the group $SU(1,1)$. The last requirement is
imposed by the fact that the solution of eq.(\ref{pic}) has to be
single  valued. 
Eq.(\ref{pic}) is the type of equation one encounters in the ADM
treatment \cite{MS,CMS} of $2+1$ gravity coupled with point
particles in the maximally slicing gauge \cite{BCV}. In this
case $z$ varies on the Riemann sphere, ${\cal N}$ of the $z_j$ are the
particle singularities with 
residue $g_j = -1 +\mu_j$ and ${\cal N}-2$ of them are the so called
apparent singularities $z_B$ with residues $g_B=1$. The inequalities
on the values of $g_m$ are satisfied in $2+1$ dimensional gravity due
to the restriction on the masses of the particles $0<\mu_n<1$ (in
rationalized Planck units) and to the fact that the total energy $\mu$
must satisfy the bound $\sum_n \mu_n <\mu<1$. 
For this reason in this paper we
shall confine ourselves to the Riemann sphere.
After solving all the
constraints, the hamiltonian nature of the particle equations of
motion is a consequence of Polyakov conjecture; actually is the
consequence of a somewhat weaker form of it \cite{CMS} i.e. of the
relation one obtains by taking the derivative of Polyakov conjecture
with respect to the total energy.

From eq.(\ref{pic}) one can easily prove
\cite{poincare,bilalgervais} that the function $Q(z)$ defined by
\begin{equation}\label{bg}
e^{\frac{\phi}{2}} \partial_z^2 e^{-\frac{\phi}{2}} = -Q(z)
\end{equation}
is analytic i.e. as pointed out in \cite{bilalgervais} $Q(z)$ is given by the
analytic component of the energy momentum tensor of a Liouville
theory. $Q(z)$ is meromorphic with poles
up to the second order \cite{hempel} i.e. of the form 
\begin{equation}\label{Q}
Q(z) = \sum_n - \frac{g_n(g_n+2)}{4(z-z_n)^2} +
\frac{\beta_n}{2(z-z_n)}. 
\end{equation}
All  solutions of eq.(\ref{pic}) can be put in the form
\begin{equation}\label{mapping}
e^{\phi}=\frac{8f'\bar{f'}}{(1-f\bar f)^2} = \frac{8
|w_{12}|^2}{(y_2\bar y_2 - y_1\bar y_1)^2},~~~~
f(z) = \frac{y_1}{y_2}
\end{equation}
being $y_1,y_2$ two properly chosen, linearly independent solutions
of the fuchsian equation 
\begin{equation}\label{fuchs}
y''+Q(z)y=0.
\end{equation}
$w_{12}$ is the constant wronskian. 
In fact following
\cite{poincare,bilalgervais} as $e^{-\phi/2}$ solves the fuchsian
equation (\ref{fuchs})
it can be put in the form
\begin{equation}\label{bgform1}
e^{-\frac{\phi}{2}}=\frac{1}{\sqrt{8}}[\psi_2(z)\bar\chi_2(\bar z) -
\psi_1(z)\bar \chi_1(\bar z)]
\end{equation}
with $\psi_j(z)$ solutions of eq.(\ref{fuchs}) with wronskian
$1$ and $\chi_j(z)$ also solutions of eq.(\ref{fuchs}) with
wronskian $1$.   
The solution of eq.(\ref{pic}) ($\phi={\rm real}$)  with the stated
behavior at infinity is unique
\cite{picard,troyanov}. Exploiting the reality of $e^{\phi}$ it is
possible by an $SL(2C)$ transformation to reduce eq.(\ref{bgform1}) to
the form 
eq.(\ref{mapping}). In fact, being $\chi_j$ linear combinations of the
$\psi_j$, the reality of $e^{\phi}$ imposes 
\begin{equation}\label{bgform2}
\psi_2(z)\bar\chi_2(\bar z) - \psi_1(z)\bar \chi_1(\bar z)=
\sum_{jk} \bar\psi_j H_{jk}\psi_k
\end{equation}  
with the $2\times 2$ matrix $H_{jk}$  hermitean and $\det H = -1$. By
means of a unitary 
transformation, which belongs to $SL(2C)$ we can reduce $H$ to
diagonal form ${\rm diag}(-\lambda, \lambda^{-1})$ and with a
subsequent $SL(2C)$ transformation we can reduce it to the form ${\rm
diag}(-1,1)$ i.e. to the form (\ref{mapping}). 
Through eq.(\ref{bg}) $\phi$ contains the full information about the
accessory parameters $\beta_n$ defined in eq.(\ref{Q}). 
It is important to
notice that being all of our monodromies elliptic, we can by means of
an $SU(1,1)$ transformation, choose around a given singularity $z_m$ (not
around all singularities simultaneously) $y_1$ and $y_2$ with the following
canonical behavior
\begin{equation}\label{canonical}
y_1(\zeta) = k_m \zeta^{\frac{g_m}{2}+1} A(\zeta),~~~~y_2(\zeta)
=\zeta^{-\frac{g_m}{2}}  B(\zeta)  
\end{equation}  
with $\zeta= z-z_m$ and $A$ and $B$ analytic functions of $\zeta$ in a
neighborhood of $0$ with $A(0)=B(0)=1$.   

%
%

\section{The realization of fuchsian SU(1,1) monodromies}

The result of Picard assures us that given the position of the
singularities $z_n$ and the classes of monodromies characterized by
the real numbers $g_n$ 
there exists a unique fuchsian equation which realizes $SU(1,1)$
monodromies of the prescribed classes. In particular the uniqueness of
the solution of Picard's equation tells us that the accessory
parameters $\beta_i$ are single valued functions of the parameter $z_n$
and $g_n$. We shall examine in this section how such dependence arises
from the viewpoint of the imposition of the $SU(1,1)$ condition on the
monodromies in order to understand the nature of the dependence of the
$\beta_i$ on the $g_n$ and on the $z_n$.  The proof of the real
analytic dependence of the accessory parameters on the $z_n$ in the
case of rational $g_n$ has been given by Kra \cite{kra}.

Starting from the singularity in $z_1$
we can consider the canonical pair of solutions around $z_1$
i.e. those solutions which behave as a single fractional power
multiplied by an analytic function with coefficient one as given in
eq.(\ref{canonical}).  We shall call 
such pair of 
solutions $(y^1_1, y^1_2)$ and let $(y_1, y_2)$ the solution which
realize $SU(1,1)$ around all singularities. Obviously all conjugations
with any element of $SU(1,1)$ is still an equivalent solution in the
sense that they provide the same conformal factor $\phi$.
The canonical pairs around different singularities are linearly
related i.e. $(y^1_1, y^1_2) = (y^2_1, y^2_2) C_{21}$.
We fix the conjugation class by setting
\begin{equation}
(y_1, y_2) = (y^1_1, y^1_2) K
\end{equation}
with $K = {\rm diag}( k, k^{-1})$ being the overall constant irrelevant
in determining $\phi$. Moreover if the solution $(y_1, y_2)$ realizes
$SU(1,1)$ monodromies around all singularities also $(y_1,
y_2)\times{\rm diag}(e^{i\alpha},e^{-i\alpha})$ accomplishes the same purpose
being ${\rm diag}(e^{i\alpha},e^{-i\alpha})$ an element of
$SU(1,1)$. Thus the phase of the number $k$ is irrelevant
and so we can consider it real and positive. This choice of the canonical
pairs is always possible in our case. In fact the roots of the indicial
equation are $-\frac{g_m}{2}$ and $\frac{g_m}{2}+1$
and thus the monodromy matrix has eigenvalues $e^{-i\pi g_m}$ and $e^{i\pi
g_m}$ which are different when $g_m$ is not an integer. If $g_m$ is an
integer in general in the solution of the fuchsian equation the less
singular solution possesses a logarithmic term which however has to be
absent in our case (no logarithm condition \cite{yoshida}) in order to
have a single valued $\phi$. In this case the monodromy
matrix is simply the identity or minus the identity.
The monodromy around $z_1$ thus belongs to $SU(1,1)$ for any choice of
$K$. 
If $D_{n}$ denote the diagonal monodromy matrices around $z_n$, we
have that the monodromy around $z_1$ is $D_1$ and the one around $z_2$
is
\begin{equation}
M_2 = K^{-1} C_{12} D_2 C_{21} K.   
\end{equation}
where with $C_{12}$ we have denoted the inverse of the $2\times 2$
matrix $C_{21}$. 

In the case of three singularities (one of them at infinity) the
counting of the 
degrees of freedom is the following: by using the freedom on $K$ we
can reduce $M_2$ to the form 
$\begin{pmatrix}
a & b \\ c & d
\end{pmatrix} 
$
with ${\rm Re}~ b ={\rm Re}~ c$, or if either  ${\rm Re}~ b$ or ${\rm
Re}~c$ is zero we can 
obtain ${\rm Im}~ b = -{\rm Im}~ c$. Then we use the fact that $D_1 M_2
= C D_\infty C^{-1}$ and thus in addition to $a+d= {\rm 
real}$ we have also $ a e^{i\pi g_1} + d e^{-i\pi g_1} =
{\rm real}$, which gives $ d= \bar a$ and thus using $a \bar a - b c
=1$ we have $c=\bar b$. The fact that a real $k$ is sufficient to
perform the described reduction of the matrix $M_2$ is assured by
Picard's result on the solubility of the problem and in this simple
case also by the explicit solution in terms of hypergeometric
functions \cite{BCV,MS}.

We give now a qualitative discussion of the
case with four singularities and then give the analytic treatment of
it. The case with more that four singularities is a trivial extension
of the four singularity case. The following treatment relies heavily
on Picard's result about the existence and uniqueness of the solution of
eq.(\ref{pic}). 
We recall that the accessory parameters $\beta_n$ are bound by two
algebraic relations known as Fuchs relations \cite{yoshida}. Thus
after choosing 
$M_1$ of the form $M_1=D_1 K$,  in imposing the $SU(1,1)$ nature of
the remaining monodromies we have at our disposal three real parameters
i.e. $k$, ${\rm Re}~\beta_3$ and ${\rm Im}~\beta_3$. It is sufficient
to impose the $SU(1,1)$ nature of $M_2$ and $M_3$ as the $SU(1,1)$
nature of $M_\infty$ is a consequence of them.  
As the matrices $M_n= K^{-1}C_{1n}D_nC_{n1}K$ satisfy identically
$\det M_n=1$ and ${\rm Tr} M_n = 2\cos\pi g_n$ we need to impose
generically on $M_2$ only two real conditions e.g. ${\rm Re}~b_2={\rm
Re}~c_2$ and ${\rm Im}~b_2=-{\rm Im}~c_2$. The same for $M_3$. Thus is
appears that we need to satisfy four real relations when we can vary
only three real parameters.  The reason why we need only three and not
four is that for any solution of the fuchsian problem the following
relation among the monodromy matrices is identically satisfied
\begin{equation}
D_1 K M_2 M_3 M_\infty=1.
\end{equation}
Rigorously the conditions for realizing $SU(1,1)$ monodromies are
\begin{equation}\label{ReIm}
{\rm Re}~a_i= {\rm Re}~d_i,~~{\rm Im}~a_i= -{\rm Im}~d_i,~~
{\rm Re}~b_i= {\rm Re}~c_i,~~{\rm Im}~b_3= -{\rm Im}~c_i~~~(i=2,3)
\end{equation}
Satisfying the eight above equations is a sufficient (and necessary)
condition to realize the $SU(1,1)$ monodromies. The fact that given a
$z^0_n$ in a neighborhood of such a point there exists one and only
one solution to the eight equation (\ref{ReIm}) means that at least
three of them 
are not identically satisfied in such a neighborhood and that the
remaining are satisfied as a consequence of them. We shall denote such
equations as
\begin{equation}\label{Deltaequations} 
\Delta^{(1)}=0,~~~~\Delta^{(2)}=0,~~~~\Delta^{(3)}=0.
\end{equation} 
The matrices $A_n= C_{n1}K$ which give the solution of the problem in
terms of 
the canonical solutions around the singularities are completely
determined by the 
two equations
\begin{equation}
(y_1,y_2) = (y^{(n)}_1,y^{(n)}_2) A_n; ~~~~
(y'_1,y'_2) = (y^{(n)'}_1,y^{(n)'}_2) A_n
\end{equation}
due to the non vanishing of the wronskian of $y^{(n)}_1,y^{(n)}_2$.
Being $(y_1,y_2)$ solutions of a fuchsian equation, $A_n$ depend
analytically  on $k, z_1, z_2, z_3 ,\beta_3$. Thus
eqs.(\ref{Deltaequations}) which
determine implicitly $k, {\rm Re}\beta_3$ and ${\rm Im}\beta_3$ state
the vanishing of the real part of three analytic functions, functions
of $z_n$, $k$ and $\beta_3$. 
It follows that $\Delta^{(i)}$ are  analytic
functions of the real and imaginary parts of the
variables or equivalently of the independent variables
$k,\beta_3,\bar\beta_3,z_n,\bar z_n$.

In order to understand the dependence of $\beta_3$ and $\bar\beta_3$
on $z_n,\bar z_n$ we apply around a solution (which due to Picard 
we know to exist) of the three equations, Weierstrass preparation
theorem \cite{gunningrossi}. It states 
that in a neighborhood of  
a solution $z_n^0$ $k^0$ $\beta_3^0$, $\Delta^{(i)}$ can be written
as
\begin{equation}
\Delta^{(i)} = P^{(i)}(k) u^{(i)}(k,z_n,\bar z_n,\beta_3,\bar\beta_3) 
\end{equation}
where $P(k)$ is a polynomial in $k$ with coefficients analytic
functions of  $z_n,\bar z_n,\beta_3, \bar\beta_3$, while 
$u(k,z_n,\bar z_n, \beta_3,\bar\beta_3)$ is a ``unit''   
i.e. an analytic function of the arguments, which does not vanish in a
neighborhood of $z_n^0, k^0, \beta^0$. Thus our problem is reduced to
the search of the real common zeros of the three polynomial
$P^{(i)}(k)$. By algebraic elimination of the variable $k$ we reach a
system of two equations which depend analytically on the variables
$z_n,\bar z_n,\beta_3,\bar \beta_3$ and reasoning as above by
elimination of $\bar \beta_3$ we reach as condition to be satisfied
by Picard's solution
\begin{equation}
V(\beta_3| c_k(z_n,\bar z_n))=0
\end{equation}
where $V$ is a polynomial in $\beta_3$ with coefficients
analytic functions of $z_n$ and $\bar z_n$. The derivative
of $\beta_3$ with respect to $z_n$ (and similarly with respect to
$\bar z_n$) is then given by 
\begin{equation}
V'(\beta_3|c_k)\frac{\partial\beta_3 }{\partial z_n} +
V(\beta_3|\frac{\partial c_k}{\partial z_n})=0  
\end{equation}
If $V'$ vanishes identically on the $\beta_3(z_n)$ provided
by Picard's solution we can adopt $V'$ as determining such a
function. The procedure can be repeated until $V'$ does not vanish
identically on  Picard's solution and thus in a neighborhood of
$z^0_n$ the derivative $\displaystyle{\frac{\partial\beta_3}{\partial
z_n}}$   
exists except for a finite number of points. Actually $\beta_3$ is an
analytic functions of $z_n$ and $\bar z_n$ for all points of such a
neighborhood in which $V'$ does not vanish \cite{gunningrossi}.
The extension to five or more singularities proceeds along the same
line.

As we already mentioned if some of the $g_m$ is an integer we have the
so called apparent singularities which have monodromy $I$ if $g_m$ is
even  and monodromy $-I$ if $g_m$ is odd. In this case we have to
impose the so called no-logarithm conditions (see e.g. \cite{yoshida})
which result in a linear
combination of the $\beta_n$ with $n\neq m$ to be equal to the square
of $\beta_m$. Thus we can eliminate one $\beta_n$ in favor of
$\beta_m$ and we have the same matching in the degrees of freedom.

\section{Proof of Polyakov conjecture}\label{direct}

As already stated we shall limit ourselves to the case of the Riemann
sphere with a finite number of conical singularities, one of them at
infinity, subject to the restrictions given by Picard and described in
sect.1. The technique to 
prove Polyakov conjecture will be to express the 
original action in terms of a 
field $\phi_M$ which is less singular than the original conformal
field $\phi$. This procedure will 
give rise to an action $S$ for the field $\phi_M$ which does not
involve the $\epsilon\rightarrow 0$ process.
Despite that, computing the derivative of the new action $S$ 
is not completely trivial because one cannot take directly the
derivative operation under the integral sign. In fact such unwarranted
procedure would give rise to an integrand which is not absolutely
summable. 
In the global coordinate system $z$ on $C$  one writes $\phi = \phi_M
+\phi_0+\phi_B$ 
where $\phi_B$ is a background conformal factor which is regular and
behaves at infinity  like $\phi_B = -2\ln(z\bar z)+c_B+O(1/|z|)$ while
$\phi_0$ is a 
solution of  
\begin{equation}\label{eqphi0}
4 \partial_z\partial_{\bar z}\phi_0 = 4\pi\sum_n g_n \delta^2(z-z_n)
-4 \alpha \partial_z\partial_{\bar z}\phi_B 
\end{equation}
with behavior at infinity $\phi_0 = (2-g_\infty)\ln(z\bar z) +
O(1)$. Such a behavior fixes the value of $\alpha$ to 
\begin{equation}\label{alpha}
4\pi (\sum_n g_n  +g_\infty -2)+8\pi\alpha=0  
\end{equation}
(a possible choice for $\phi_B$ is the conformal factor of the sphere
i.e. $\phi_B = -2\ln(1+z\bar z)$).
The fields $\phi_0$ and $\phi_M$ transform under a change of chart
like scalars while $e^{\phi_B}$ transforms as a $(1,1)$ density. This
choice is also in agreement with the invariance of
eq.(\ref{eqphi0},\ref{eqphiM},\ref{Saction}). 
The expression of $\phi_0$
is
\begin{equation}
\phi_0 = \sum_n g_n \ln|z-z_n|^2 
-\alpha \phi_B+c_0.
\end{equation}
Then
we have for $\phi_M$
\begin{equation}\label{eqphiM}
4\partial_z\partial_{\bar z}\phi_M =
e^{\phi_0+\phi_B+\phi_M}+(\alpha-1)4~\partial_z\partial_{\bar z} \phi_B.  
\end{equation}
$\phi_M$ is a continuous function on the Riemann sphere.
The action which generates the above equation is
\begin{equation}\label{Saction}
S=\int d\mu [e^{-\phi_B}
\partial_z\phi_M \partial_{\bar z}\phi_M + \frac{e^{\phi_0+\phi_M}}{2}+
2(\alpha-1)e^{-\phi_B}\phi_M\partial_z\partial_{\bar z}\phi_B]=
\end{equation}
$$ 
\int [
\partial_z\phi_M \partial_{\bar z}\phi_M + \frac{e^{\phi}}{2}+
2(\alpha-1)\phi_M\partial_z\partial_{\bar z}\phi_B] \frac{i dz\wedge
d\bar z}{2} \equiv \int d\mu ~F~~~{\rm with}~~~
d\mu \equiv e^{\phi_B} \frac{i dz\wedge d\bar z}{2} 
$$
where the splitting between the measure and the integrand
has been introduced for later convenience. Due to the behavior of
$\phi_M$ and $\phi_0$ at the singularities and at infinity the
integral in eq.(\ref{Saction}) converges absolutely. It is
straightforward to prove that  the action $S$ computed on the solution
of eq.(\ref{eqphiM}) is related to the original Polyakov action $S_P$
also computed on the solution of eq.(\ref{eqphiM}) by
$$
S_P = S - (\alpha-1)^2\int \phi_B\partial_z\partial_{\bar z}\phi_B
\frac{idz\wedge d\bar z}{2} + 2\pi (\alpha -1)^2c_B+
$$
\begin{equation}\label{SPtoS}
+\pi\sum_m\sum_{n\neq m} g_m g_n \ln|z_m-z_n|^2 
+4\pi c_0(1-\alpha).
\end{equation}
The behavior of $\phi_M$ around the singularities $z_m$ can be deduced 
from 
eqs.(\ref{mapping},\ref{canonical}).
Thus in a finite neighborhood of $z_m$ we can write
\begin{equation}\label{expansionM}
\phi_M = \sum_{nLMN} c_{nLMN} [(z-z_n) (\bar z -\bar
z_n)]^{L(g_n+1)} (z-z_n)^M (\bar z-\bar z_n)^N \rho(|z-z_n|)+\phi_{Mr}   
%
\end{equation}
where the finite sum extends to the terms such that $2L(g_n+1)+M+N\leq
3$ and $\rho(|\zeta|)$ is chosen $C^\infty$ with $\rho=1$  in a finite
neighborhood of $0$  and zero for $|\zeta|>1$.   
$\phi_{Mr}$
%
is a continuous function $O(|z-z_n|^3)$ around each $z_n$ .
We saw in
sect.2 how except for a finite number of points in a neighborhood of
$z_n$ there exists the derivative of the parameters $k,{\rm
Re}\beta_i,{\rm Im}\beta_i$ which determine the solutions of the
fuchsian equation related by eq.(\ref{mapping}) to the conformal factor
$\phi$. Actually as pointed out at the end of sect.(2) around
the points where $V'$ does
not vanish such parameters are analytic functions of $z_n$. On the
other hand the solutions of the fuchsian equation and 
thus $\phi_M$ depend analytically on such parameters \cite{hille}.

The procedure to compute the derivative will be to prove that 
\begin{equation}\label{limit}
\frac{\partial S}{\partial z_m}= \lim_{\epsilon \rightarrow 0}
\int_{X_\epsilon} d\mu \frac{\partial F}{\partial z_m} 
\end{equation}
where $F$ is given in eq.(\ref{Saction})
and $X_\epsilon$ has been defined after eq.(\ref{Sepsilon}).

This is achieved by writing $F=(F-f) + f$
where $F-f$ is sufficiently regular, i.e. continuous and absolutely
integrable with
$\frac{\partial (F-f)}{\partial z_m}$ continuous and
$|\frac{\partial (F-f)}{\partial z_m}| < M$  for any $z$ while $z_m$
varies in a finite interval, so that   
\begin{equation}\label{limit1}
\frac{\partial}{\partial z_m}\int (F-f) d\mu=\int\frac{\partial}{\partial
z_m}(F-f) d\mu \equiv\lim_{\epsilon \rightarrow
0}\int_{X_\epsilon}\frac{\partial}{\partial z_m}(F-f) d\mu   
\end{equation}
proving at the same time that
\begin{equation}\label{limit2}
\frac{\partial}{\partial z_m}\int f d\mu =\lim_{\epsilon \rightarrow
0}\int_{X_\epsilon}\frac{\partial}{\partial z_m} f d\mu.   
\end{equation}
Then by summing eq.(\ref{limit1}) and eq.(\ref{limit2}) we obtain
eq.(\ref{limit}). 

Using the expansions eq.(\ref{expansionM}) we shall
choose the function $f$ as 
\begin{equation}
f d\mu = \sum_{nLMN} b_{nLMN}\frac{[(z-z_n)(\bar z- \bar
z_n)]^{L(g_n+1)}}{(z-z_n)(\bar z - \bar z_n)}(z-z_n)^M(\bar z-\bar z_n)^N
\rho(|z-z_n|) \frac{i dz\wedge d\bar z}{2}
\end{equation}
$$
\equiv\sum_{nLMN} b_{nLMN}  G_{nLMN}(z-z_n) \frac{i dz\wedge d\bar z}{2}
$$ 
where the finite sum extend to all singularities of $F$ and $M,N\geq
0$ and $L\geq 1$ such that $2L(g_m+1)+M+N\leq 3$. We notice that  
\begin{equation}
\frac{\partial}{\partial z_m}\int f d\mu =\frac{\partial}{\partial
z_m}\sum_{nLMN} b_{nLMN}\int G_{nLMN}(z-z_n) 
\frac{i dz\wedge d\bar z}{2} =
\end{equation}
$$
\sum_{nLMN} \frac{\partial b_{nLMN}}{\partial z_m}
\int G_{nLMN}(z-z_n) 
\frac{i dz\wedge d\bar z}{2} 
$$
the point being that each integral in the sum does not depend on
$z_n$ due to translational invariance and thus we have to take only the
derivative of the coefficients $b_{nLMN}$. Moreover
$$
\int_{X_\epsilon}\frac{\partial f}{\partial z_m}d\mu =
\sum_{nLMN} \frac{\partial b_{nLMN}}{\partial z_m}
\int_{X_\epsilon} G_{nLMN}(z-z_n) 
\frac{i dz\wedge d\bar z}{2} 
$$
\begin{equation}
-\sum_{LMN} b_{mLMN}
\int_{X_\epsilon}\frac{\partial}{\partial z} G_{mLMN}(z-z_m) 
\frac{i dz\wedge d\bar z}{2}.
\end{equation}
The last term is either zero due to the phase of the integrand or goes
to zero for $\epsilon\rightarrow 0$ by power counting and thus we have
the stated result eq.(\ref{limit2}). 
We are left to prove that in eq.(\ref{limit1}) we can take the
derivative operation under the integral sign. To this purpose it is
sufficient to prove that $F-f$ and $\frac{\partial (F-f)}{\partial
z_n}$ 
are continuous on the product of the Riemann sphere and a closed disk
of $z_n$ having for center the values of $z_n$ for which according to
sect.2 the derivative of $k,\beta_n ,\bar\beta_n$
exists. In fact $F-f$ and $\frac{\partial (F-f)}{\partial z_n}$ are
free of singularities both at the finite and at infinity. As the product of
the Riemann sphere and a closed disk is a compact set the hypothesis
above stated are satisfied and this allows the exchange of the
derivative operation with the integral sign.
Using now the equation of motion (\ref{eqphiM}) we obtain 
\begin{equation}
\frac{\partial S}{\partial z_m} = \lim_{\epsilon
\rightarrow 0} 
\int_{X_\epsilon}[\partial_z(\frac{\partial \phi_M}{\partial
z_m}\partial_{\bar z} \phi_M) +
\partial_{\bar z}(\frac{\partial \phi_M}{\partial
z_m}\partial_{z} \phi_M) 
+\frac{\partial \phi_0}{\partial
z_m}\frac{e^\phi}{2}]\frac{i dz\wedge d\bar z}{2}. 
\end{equation}
It is easily checked that the only contribution which survives in the
limit $\epsilon \rightarrow 0$ is
\begin{equation}
\frac{\partial S}{\partial z_m} = \lim_{\epsilon
\rightarrow 0}
\int_{X_\epsilon} \frac{e^\phi}{2}\frac{\partial \phi_0}{\partial z_m} 
\frac{i dz\wedge d\bar z}{2}
\end{equation}
which can be computed by using eq.(\ref{eqphiM}) and
$\displaystyle{\frac{\partial \phi_0}{\partial z_m} = - \frac{g_m}{(z-z_m)}}$
to obtain
\begin{equation}\label{contour}
\frac{\partial S}{\partial z_m} = - i g_m \lim_{\epsilon \rightarrow 0}
\oint_{\gamma_{\epsilon}} \frac{1}{z-z_m}
\partial_z\left(\phi_M-(\alpha-1)\phi_B\right) d z.
\end{equation}
Using $\phi_M -(\alpha-1)\phi_B = \phi -\sum_n g_n\ln|z-z_n|^2$ and
the expansion of 
$A=1+c_1\zeta+\cdots$ and $B
=1+ c_2\zeta + \cdots$ which are obtained by substituting into the
differential equation (\ref{fuchs}) to obtain
\begin{equation}
c_1=-\frac{\beta_m}{2(g_m+2)}~~~~{\rm and}~~~~c_2= \frac{\beta_m}{2g_m}
\end{equation}
finally we have
\begin{equation}
\frac{\partial S}{\partial z_m} = 
-2\pi \beta_m - 2\pi\sum_{n,n\neq m} \frac{g_m g_n}{z_m-z_n} 
\end{equation}
equivalent to Polyakov conjecture eq.(\ref{conjecture}) due to the
relation (\ref{SPtoS}) between $S$ and $S_P$.

\section*{Acknowledgments}
We are grateful to Mauro Carfora for pointing out to us reference
\cite{troyanov} and for useful discussions.

\end{document}